\documentclass[conference]{IEEEtran}
\IEEEoverridecommandlockouts
\usepackage{cite}
\usepackage{amsmath,amssymb,amsfonts}
\usepackage{algorithmic}
\usepackage{graphicx}
\usepackage{textcomp}
\usepackage{xcolor}
\usepackage{booktabs}
\usepackage{threeparttable}
\usepackage{enumitem}
\usepackage{hyperref}
\usepackage{listings}
\usepackage{subcaption}

\def\BibTeX{{\rm B\kern-.05em{\sc i\kern-.025em b}\kern-.08em
    T\kern-.1667em\lower.7ex\hbox{E}\kern-.125emX}}
\begin{document}

\title{Demystifying AMD SEV Performance Penalty for NFV Deployment}

\author{\IEEEauthorblockN{Syafiq Al Atiiq}
\IEEEauthorblockA{\textit{Dept. of Electrical and Information Technology} \\
\textit{Lund University}, Sweden\\
syafiq\_al.atiiq@eit.lth.se}
\and
\IEEEauthorblockN{Aris Cahyadi Risdianto}
\IEEEauthorblockA{\textit{Information and Communication Technology Cluster} \\
\textit{Singapore Institute of Technology (SIT), Singapore}\\
ariscahyadi.risdianto@singaporetech.edu.sg}
\and
}

\maketitle

\begin{abstract}
Network Function Virtualization (NFV) has shifted communication networks towards more adaptable software solutions, but this transition raises new security concerns, particularly in public cloud deployments. While Intel's Software Guard Extensions (SGX) offers a potential remedy, it requires complex application adaptations. This paper investigates AMD's Secure Encrypted Virtualization (SEV) as an alternative approach for securing NFV. SEV encrypts virtual machine (VM) memory, protecting it from threats including those at the hypervisor level, without requiring application modifications. We explore the practicality and performance implications of executing native network function (NF) implementations in AMD SEV-SNP, the latest iteration of SEV. Our study focuses on running an unmodified Snort NF within SEV. Results show an average performance penalty of approximately 20\% across various traffic and packet configurations, demonstrating a trade-off between security and performance that may be acceptable for many NFV deployments.
\end{abstract}


\section{Introduction}
In recent years, the networking domain has experienced a significant transformation with the advent of Software Defined Networking (SDN) and Network Function Virtualization (NFV). These technologies have shifted the focus from traditional hardware-centric operations to software-based models. While this shift has introduced several advantages, such as enhanced scalability and cost-effectiveness, it has concurrently posed new security and privacy challenges. One of the most important challenges is the trustworthiness of cloud service providers and their co-tenants. Traditionally, security frameworks have regarded hypervisors as trust anchors due to their minimal interface and limited direct interaction with applications. Their role, primarily to manage VMs and resources, was perceived to have a reduced attack surface. However, with a series of vulnerabilities\footnote{https://nvd.nist.gov/vuln/detail/CVE-2019-19332}\footnote{https://nvd.nist.gov/vuln/detail/CVE-2019-14821} uncovered in recent years, the community began to realize that even hypervisors might not be immune to attacks. Additionally, as the number of services and applications relying on virtualized environments grew, the stakes related to hypervisor security escalated. Furthermore, there's a growing sentiment, underlined by recent research \cite{boneh2015hosting}, suggesting that for certain applications or industries, it might be more prudent to treat hypervisors and adjacent VMs as potential threat vectors.

Previously, an approach to address these concerns has been the use of Intel SGX \cite{10.1145/3040992.3040994,snort-sgx,9039980,10.1145/2876019.2876032,10.1145/3180465.3180476}. However, we see at least three reasons why such solutions do not live up to the expectations. First, SGX often requires significant application adaptation to fit within the Trusted Execution Environment (TEE) \cite{8548184}, and the limited resources available within such environments can pose substantial challenges, particularly for NF with high resource requirements. Second, over recent years, SGX has faced significant challenges due to a multitude of issues \cite{217543,sgaxe,schaik2021cacheout,van2022sok}. Intel has diligently worked to enhance SGX's security by releasing patches and introducing new (micro)architectures. However, this continuous evolution makes it progressively more challenging to assess the effectiveness of different attack methods in SGX. Lastly, SGX is not designed to perform VM isolation, is used to protect application from a potentially compromised OS. In contrast, VM security is often about isolating entire OS instances from one another and from the hypervisor. These are different threat models and require different mechanisms to address.

In the subsequent developments, Intel has introduced Trust Domain Extensions (TDX) \cite{Johnson2023}. However, at the time of writing this paper, no source code has been released to the public to accompany the provided documentation; hence, there is no opportunity for us to scrutinize the performance. At the same time, AMD has already released its SEV solution, employing unique keys for each VM, ensuring isolation between guest-to-guest and guest-to-hypervisor interactions. SEV enhances its protective capabilities by allowing scenarios where the hypervisor and neighbouring VM are considered untrusted entities. This trust model enables a trusted VM to coexist securely with other untrusted elements. Unlike SGX, which requires stack adaptation for NFs applications, SEV treats the VM itself as a trusted entity, eliminating the need for such modifications \cite{snort-sgx}. This streamlined approach significantly simplifies the deployment of secure NFs in cloud environments, facilitating a more efficient process for securing NFVs. Drawing upon previous research on NF's deployment with SGX \cite{snort-sgx,9039980,10.1145/2876019.2876032,10.1145/3040992.3040994,10.1145/3180465.3180476}, this paper presents firsthand experiences in running an intrusion detection system (IDS), Snort in particular, on top of AMD SEV. In summary, here are our contributions:

\begin{enumerate}
    \item Verify whether Snort's original code base can run in a trusted VM in SEV without any changes or additional implementation required.
    \item Measure and analyze the performance and the adaptation effort difference between Snort's deployment over AMD SEV and Intel SGX.
\end{enumerate}

We follow the pattern of a previous study, SEC-IDS \cite{snort-sgx} on Intel SGX, where Snort is used to measure the performance when it is placed inside the enclave. Though our comparison is not direct (since we integrate Snort within a VM, while the prior study incorporated it into a baremetal environment), it provides a reference for system administrators considering the deployment of applications in a trusted VM. The measurement in our paper can be treated as ``the additional cost of security'' when one decides to put the application inside the trusted VM. SEV-SNP, the latest release of SEV, is used in our measurement. We call our contribution \textbf{SEV-IDS}. 

The paper is organized as follows. We begin by providing background information on AMD SEV in section \ref{sec_background}. Then, in Section \ref{sec_threatmodel}, we describe the threat model of SEV-IDS. Section \ref{sec_architecture} explains the proof-of-concept we use in this paper. Section \ref{sec_measurement} provides the measurement and analysis of the system. We present the related work in section \ref{sec_rw}. Finally, we draw conclusions and anticipate future works in Section \ref{sec_conclusion}.

\section{Background}
\label{sec_background}

\subsection{Security Challenges for NFV}
From the security standpoint, running NF with the assumption that the hypervisor and the adjacent VM might behave maliciously is a non-trivial task. We identify three primary challenges associated with such an undertaking:

\begin{enumerate}[label=C\arabic*]
    \item \label{C1}\textbf{Secure Data Processing}. Most TEE-based NF security models require a distinction between sensitive and non-sensitive components or processes. For instance, in \cite{snort-sgx}, while Snort's binary and associated plugins/libraries are treated as sensitive, the configuration and rules are not.
    \item \label{C2}\textbf{State Protection}. A common requirement for NFs is to preserve the result or status of data processing, referred to as the \textit{state}. For states that reside in memory, such as the execution state and buffer, the protection strategy is similar to that of data processing due to the AES encryption and other safeguards, including page validation and VM privilege level differentiation \cite{Kaplan2020}. For states saved in files (e.g., rules, logs), another approach is necessary.
    \item \label{C3}\textbf{Integrity Checks}. Deploying NF in TEE-based cloud infrastructures requires two distinct integrity checks: one to confirm that the environment aligns with the NF owner's specified security level and another to ascertain that the trusted NF software is being run \cite{safebrickspoddar2018}.
\end{enumerate}

In light of these challenges, the upcoming subsections delve into the potential of AMD SEV and its fundamental component, memory encryption, as solutions. 

\subsection{Secure Memory Encryption (SME)}
\label{sec:sme}

AMD has introduced a new CPU architecture that features Secure Memory Encryption (SME) \cite{208506}, which is designed to encrypt the primary memory. This encryption process is executed by specialized hardware located in the on-chip memory controllers. Each of these controllers contains an AES engine responsible for encrypting data before it's stored in the DRAM and for decrypting data upon retrieval. For security enhancement, with every system reboot, a distinct encryption key is generated. This key remains inaccessible to the ongoing CPU processes. The management of this key is entrusted to the AMD Secure Processor (AMD-SP), a distinct ARM Cortex-A5 processor embedded within the AMD System On Chip (SOC), dedicated solely to key management functions. The decision to encrypt specific memory pages rests with the operating system or the hypervisor. This choice is signified using a dedicated physical address bit, specifically the $47^{th}$ bit, referred to as the C-bit. When the operating system wants a memory page to undergo encryption, it sets the C-bit to a value of 1. Consequently, data written to this page will be encrypted, and during retrieval, decryption is performed.

\subsection{Secure Encrypted Virtualization (SEV)}

Building on the foundation set by SME described in Section \ref{sec:sme}, AMD introduced Secure Encrypted Virtualization (SEV) \cite{Kaplan2020}. Traditional computing, governed by the ring-based security model \cite{6234805}, permits higher privileged codes to access resources of their tier and below. As platforms like the Linux kernel evolve in complexity, challenges in software integrity and resistance to potential breaches arise. AMD's SEV introduces a solution through its cryptographic mechanism, setting clear boundaries between privilege levels. This design aims to strengthen the security of lower-level entities without an implicit trust in higher-tier software. SEV's approach changes how we view threats compared to the traditional ring-based model. It suggests that an attacker might work within low-privilege areas like VMs and simultaneously access more secure sections, like the hypervisor. SEV's capabilities extend beyond isolating large software entities, i.e., full-fledged VMs with dedicated resources. It can also be applied for the protection of smaller software units, including containers. However, in our work, we limit the measurement only to the VM case.

At its core, SEV integrates the security of SME features with AMD's established AMD-V virtualization architecture. This provides an environment where the entire VM can operate in an encrypted state. AMD suggests that the latency overhead introduced due to the continuous encryption and decryption during DRAM access is negligible. This claim has been supported by empirical studies \cite{Larabel2022}. 

AMD releases SEV in three stages to cover different attacks that we will explain them based on threat model's perspective:

\subsubsection{SEV}
Initial SEV release covers attack vectors where the adversary can read the VM memory. Such an attack can be launched either from the hypervisor (host machine) or from the Direct Memory Access (DMA) devices. This version also covers the denial-of-service attack coming from the malicious guest towards the hypervisor, i.e., a guest who refuses to yield/exit. Lastly, it covers an offline DRAM analysis (e.g., cold boot attack) to analyze the content after power loss. 

\subsubsection{SEV-ES}
In the second iteration of SEV, called Encrypted State (SEV-ES), an additional safeguard was introduced to harden the CPU register state. Despite the SEV feature being enabled, an adversarial hypervisor could potentially steal this information from the registers or manipulate the guest state, including encryption keys, pointers, and more. To mitigate such attacks, SEV-ES encrypts the contents of CPU registers following a VMEXIT event, which occurs when the VM terminates the execution \cite{Kaplan2017}.

\subsubsection{SEV-SNP}
SEV's third variant, SEV-SNP (Secure Nested Paging) \cite{Kaplan2020}, enhanced VM isolation with integrity protection features (in addition to the previously covered threats). Previous versions could not shield against a malicious hypervisor altering the guest operating system's page tables, which posed risks of unauthorized data exposure or malicious code insertion. SEV-SNP addresses four key integrity threats:

\begin{itemize}
    \item Replay Protection and Data Corruption - Prevents untrusted entities from writing to protected VM memory by tracking ownership of memory pages using the Reverse Map Table (RMP).
    \item Memory Aliasing - Avoids unintended data corruption using RMP for memory page one-to-one mapping when counters hypervisor attempts to map two guest pages to one physical memory page.
    \item Memory Re-Mapping - Restricts the hypervisor from mapping a guest page to various physical memory pages by controlling it from trusted entities (i.e., AMD-SP) and pairing the new RMP with the VM code.
    \item Interrupt Handling Enhancements - SEV-SNP introduces restricted injection and alternate injection modes by permitting VMs to manage interrupt handling and Advanced Programmable Interrupt Controller (APIC) emulation.
\end{itemize}

\begin{figure}[t]
    \centering
    \includegraphics[width=0.5\linewidth]{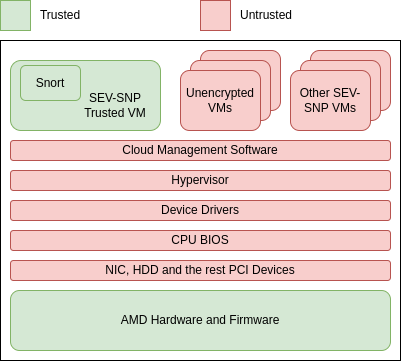}
    \caption{Threat model: A single physical machine perspective.}
    \label{fig:tm_1machine}
\end{figure}

The SEV's threat model in a single physical machine perspective is shown in figure \ref{fig:tm_1machine}, where the only trusted entities are the trusted VM itself and the SEV hardware and their respective firmware. Using this, we intend to run an IDS, snort \cite{roesch1999snort} in particular, and conclude the feasibility in regards to its performance penalty. 

\subsection{How SEV solves the challenges}
For \ref{C1}, using SEV, both the binary and the associated plugins can operate within the trusted VM, which streamlines the implementation process as it mirrors the functionality of a standard Linux OS. Furthermore, all NF thread processes and their corresponding libraries are stored in AES-encrypted memory spaces as the core features of SEV with a different set of security perimeters such as RMP check to restrict access to guest's virtual-to-physical address space owned by SEV-SNP, and Virtual Machine Privilege Level (VMPL) assignment to apply security control in VM-to-hypervisor communication \cite{Kaplan2020}. For \ref{C2}, while SEV does not offer a direct solution for this, it does recommend the use of a fully encrypted disk (FDE) for the guest VM \cite{Kaplan2020}. The choice of encryption methods can vary depending on the capabilities of the hypervisor. For example, current Linux hypervisors, such as QEMU/KVM, already support AES encryption in the QCOW2 disk format \cite{qemudiskimage2022}. 

For the first check on \ref{C3}, SEV utilizes TCB versioning of the AMD-SP firmware, which, combined with a secret, produces the Versioned Chip Endorsement Key (VCEK) via a cryptographic function. This unique private ECDSA key for every AMD chip is responsible for signing the attestation report. This ensures the NF owner can validate that specific security features are active within that cloud instance \cite{Kaplan2020}. The subsequent check on \ref{C3} takes place after the SEV launch of the NF guest. The NF guest owner can introduce a signed Identity Block (IDB) to uniquely identify the VM with its expected launch digest, consistently captured in all attestation reports \cite{Kaplan2020}. This allows the NF owner to verify that the NF operates on the intended cloud guest instance and not a potential duplicate created by the adversaries. Moreover, the same attestation mechanism can be employed to confirm the integrity of the NF's sensitive data during its transition from the NF owner to the guest VM.

\section{Threat Model}
\label{sec_threatmodel}

\begin{figure}[ht]
    \centering
    \includegraphics[width=0.7\linewidth]{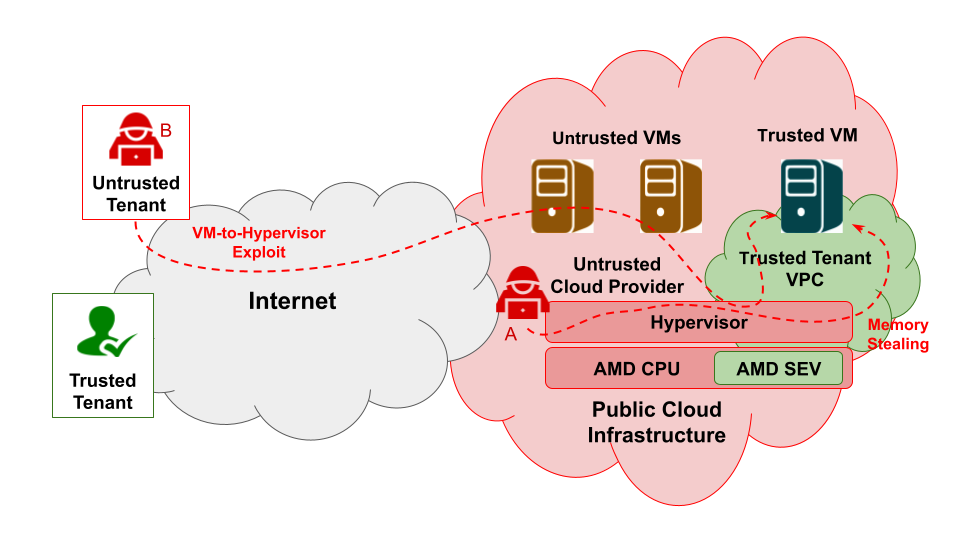}
    \caption{Threat model: A Cloud's perspective.}
    \label{fig:tm}
\end{figure}

As shown in figure \ref{fig:tm}, let $A$ be the malicious hypervisor and $B$ be the malicious tenant on the same hypervisor where SEV-IDS is being deployed. $A$ may access/alter the applied snort's rules or configuration by making an attempt to read/alter the memory and CPU registers where the operation from the trusted VM is executed. Similarly, $B$ may attempt to do the same, i.e., attempting to read/alter parts of the memory, given that there exists a vulnerability in the hypervisor that allows the untrusted VM to perform privilege escalation\footnote{https://nvd.nist.gov/vuln/detail/CVE-2022-2196}. 

\begin{table}[ht]
\centering
\caption{Threat Model}
\begin{threeparttable}
\begin{tabular}{|p{0.1\textwidth}|p{0.05\textwidth}|p{0.085\textwidth}|p{0.06\textwidth}|p{0.075\textwidth}|}
\hline
\textbf{Asset} & Integrity & Confidentiality & \multicolumn{2}{c|}{Availability, against} \\
\cline{4-5}
& & & Hypervisor & Other VMs \\
\hline
Snort execution & \checkmark & \checkmark & $\times$ & \checkmark \\
\hline
Snort state (flows, streams, metadata) & \checkmark & \checkmark\tnote{*} & $\times$ & \checkmark \\
\hline
Snort configuration and rules & \checkmark & \checkmark\tnote{*} & $\times$ & \checkmark \\
\hline
Network traffic & $\times$ & $\times$ & $\times$ & $\times$ \\
\hline
\end{tabular}
\begin{tablenotes}
\item[*] Assuming that FDE is in place.
\end{tablenotes}
\end{threeparttable}
\label{tab:tmtab}
\end{table}

It is also worth mentioning that SEV does not safeguard against availability threats or denial-of-service attacks prompted by $A$ \cite{Kaplan2020}. For instance, if $A$ refuses to run a specific guest for any reason, there is no defensive measure the guest or Snort can take. However, a denial-of-service attack aimed at the hypervisor from a guest VM sharing the same hypervisor does fall within the protective scope of SEV.

Beyond the entities previously discussed, we consider all other CPU software components—including the CPU BIOS and device drivers— and DMA devices as untrusted. In this context, being ``\textit{untrusted}'' means that we presume these components could be malicious and potentially collude with other untrusted elements to compromise the security safeguards of the trusted VM. As for Snort, we envision that all of Snort's assets are protected in terms of their integrity and confidentiality. These assets include (i) execution, (ii) state (flows, streams, metadata), and (iii) configuration and rules. It means that:
\begin{enumerate}
    \item Snort's execution is assured to run correctly, and untrusted components are unable to break the confidentiality of such execution. This is achieved through SME as the main building block of SEV.
    \item Snort's states are able to preserve their integrity and confidentiality due to the same reason as the previous item. Even when the VM is doing VMEXIT, i.e., when the execution of code within a VM is halted, and control is returned to the hypervisor.
    \item As for Snort's configuration and rules, it is trivial to make it confidential as long as the VM has FDE in place. 
\end{enumerate}

Table \ref{tab:tmtab} summarizes the threat model of SEV-IDS.

\section{Proof-of-Concept}
\label{sec_architecture}
\subsection{Implementation}
We aim to execute Snort within the trusted VM in SEV, prioritizing both minimal performance overhead and ease of deployment. Several design decisions have been made to reach this objective. Contrary to previous work \cite{snort-sgx}, which was primarily targeted at achieving near-native performance, we have chosen to abandon this ambition due to the inherent constraints associated with the VM environment. This section delves into the specifics of our design choices, outlining the considerations that helped us accomplish our objectives.

\textbf{1. The best native (non-kernel-bypass) networking.} While SEC-IDS \cite{snort-sgx} employs a Data Plane Development Kit (DPDK) to achieve the best performance, it requires dedicated cores for processing network traffic, which means those cores cannot be used for other tasks. This can lead to inefficiencies in resource usage, especially in environments where network processing is not the primary task, i.e., a virtualized environment in the cloud infrastructure. Also, using DPDK means bypassing the kernel, and one could lose out on certain kernel-provided features, like built-in security mechanisms, traffic control, and various networking tools that rely on the standard networking stack. While SEV makes it possible to treat the whole trusted VM as secure, we argue that a kernel-bypass method would negate the benefit of having rich features from the kernel. To this end, we decided to pick a native virtualized networking environment, virtio-net-pci, in KVM without kernel bypass. It is the best option for a native networking interface to have the least performance penalty in the networking stack. The driver for virtio-net-pci is paravirtualized, meaning it is aware they are running in a virtualized environment; hence, it can operate more efficiently than fully emulated devices like e1000. 

\textbf{2. Off-the-shelf and native DAQ (Data Acquisition Library).} Since version 2.9\footnote{https://www.snort.org/faq/readme-daq}, Snort separates the function call to listen to the incoming packet to a dedicated library called Data Acquisition (DAQ). For the LibDAQ library, there were three options to get the best performance out of Snort: i) AF\_PACKET, ii) Netmap, and iii) PF\_RING. We picked AF\_PACKET in the end as Netmap is also a kernel bypass (having the same argument on why we do NOT use DPDK), and PF\_RING is not natively supported by Snort. While AF\_PACKET speed does not come from bypassing the kernel networking stack, it comes from several key factors, i.e., direct access to the network interface, memory-mapped I/O (MMAP), and zero-copy mechanism. 

\subsection{On the adaptation efforts in SEC-IDS}

As we have mentioned, our intention is to run Snort securely inside a trusted VM in AMD SEV, as has been done by SEC-IDS \cite{snort-sgx}. As the whole VM is treated as secure, SEV allows us to run Snort in an off-the-shelf manner. While the following table is not intended to make an apple-to-apple comparison due to the inherent nature of VM versus baremetal, it is still an important thing to note because making Snort runs smoothly in SGX requires a lot of effort. This is not the case with AMD SEV at all. In a glimpse, table \ref{comptab} summarizes the needed adaptation to achieve the same goal of running Snort in SEV-IDS compared to SEC-IDS \cite{snort-sgx}. While SEV-IDS does not need to perform any of these adaptations, SEC-IDS requires such efforts in order to make Snort runs with the intended speed:

\begin{table}[ht]
\centering
\caption{Summary of Adaptation Efforts Needed in SEC-IDS}
\begin{tabular}{|p{0.30\textwidth}|c|c|}
\hline
\textbf{Adaptation} & \textbf{SEC-IDS} & \textbf{SEV-IDS} \\ \hline
Porting snort \& its dependencies into enclave & Yes & No \\ \hline
Packet processing outside enclave & Yes & No \\ \hline
Additional trusted clock & Yes & No \\ \hline
\end{tabular}
\label{comptab}
\end{table}

\begin{enumerate}[label=D\arabic*]
    \item \label{D1} \textbf{Porting Snort into the SGX enclave}. This posed initial challenges due to the extensive manual effort needed when using the SGX Software Development Kit (SDK). Even when all the manual effort has been made (through the help of Graphene-SGX framework \cite{203255}), modifications were needed to address issues with the hwloc, luajit, and libpcap libraries. The author ended up removing hwloc and libpcap and patching luajit from the dependency list. 
    \item \label{D2} \textbf{Packet processing outside enclave}. To avoid low-level networking support complexities, TCB bloating, and conflicts with HugePages usage, the networking stack (using DPDK) is placed outside the SGX enclave. The challenge is to facilitate communication between Snort (within the enclave) and the DPDK threads (outside the enclave). Instead of using a common ``one-thread-do-all'', the author solves the problem by separating the thread between DPDK and the Snort. This means that DPDK uses at least one thread exclusively. 
    \item \label{D3} \textbf{Trusted clock}. To measure timeouts, TCP/UDP flow expirations, packet latencies, and passage of time for statistics, Snort relies heavily on a ``clock\_gettime'' syscall. In SEC-IDS, it is initially treated as a syscall, which requires the exit of the enclave. This creates a lot of overhead as each packet will at least invoke the syscall twice. 
\end{enumerate}

For SEV-IDS, since the whole VM is treated as a secure component, there is no need to port the Snort's codebase as in \ref{D1}. While \ref{D2} might be beneficial for a system with dedicated hardware, it is arguably inefficient for a system with a shared resource, i.e., a VM in a shared hypervisor. Also, as SEV-IDS does not involve DPDK and only harnesses the standard interface from the host to the VM (virtio), no changes to the connection's mechanism are needed. And for the trusted clock in \ref{D3}, it is already inside the trusted VM, together with Snort; hence, there is no enclave exit outside the trusted VM. 

\section{Measurement and Analysis}
\label{sec_measurement}

\begin{figure}[t]
    \centering
    \includegraphics[width=0.5\linewidth]{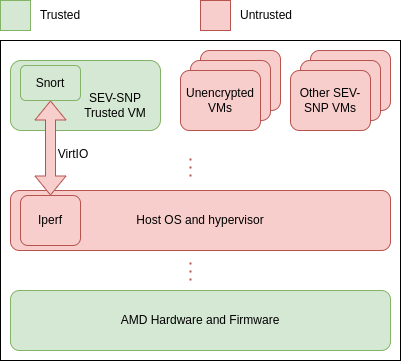}
    \caption{Measurement scenario}
    \label{fig:scen}
\end{figure}

In our experimental setup depicted in figure \ref{fig:scen}, we run Snort within the trusted VM and initiate the workload generator from the host system. While it is feasible to position the generator outside the host, we opt against this. This decision is based on the understanding that, for the trusted VM, any inbound traffic must first pass through the host system. The exception to this observation is if we employ an exclusive networking solution like DPDK. In that context, the specific location of the workload, whether inside or outside the host, becomes irrelevant. To ensure that there's no cross-interference impacting system performance, each process is specifically executed with a pre-defined CPU affinity (through a \texttt{taskset}\footnote{https://man7.org/linux/man-pages/man1/taskset.1.html}), ensuring distinct CPU cores are allocated to each process.

\subsection{Software}
We utilized the versions of software as follows: (i) Host machine: The snp-latest tree of SEV-SNP development\footnote{https://github.com/AMDESE/AMDSEV/tree/snp-latest}, (ii) trusted VM: LibDAQ v3.0.12, Snort3 v3.1.64.0, LuaJIT v2.0.5, OpenSSL v3.0.8, and PCRE 8.39. Unlike SEC-IDS, where modifications are needed in Snort, LibDAQ, and the Graphene SGX library, we did not perform any modifications to achieve the same goal. The scripts of our measurement are available at an anonymized git page\footnote{https://anonymous.4open.science/r/sev-ids-2566/README.md}.

\subsection{Hardware}
Snort runs in a trusted VM inside a server with the following specifications: AMD EPYC 7443P @ 2.85 GHz, 24 physical cores with hyper-threading enabled, RAM DDR4 64GB 3200, 2x NVME 250GB, and Ubuntu 20.04 on the host. The trusted VM is allocated with 16 GB of RAM. Due to the hyper-threading, the CPU has 48 cores, where the VM is allocated with a dedicated four cores, and the workload generator is running on a dedicated two cores.

\subsection{Methodology}
The measurement was conducted to compare SEV-IDS with the vanilla Snort. Our SEV-IDS has been described earlier, while the vanilla Snort is the same host and VM without the SEV enabled from BIOS. The workload came from iPerf2, using several different metrics, i.e., (i) packet size 128-1024 bytes and TCP flows from 1-16, and (ii) packet size 128-1024 bytes and incoming packets per second (PPS) 100-500000. While the former is generated using TCP packets, the latter is generated using UDP packets.

A measurement is taken for 2 minutes. iPerf2 initiates packet generation before Snort is activated. After the 2-minute mark, both Snort and the workload generator receive a SIGINT signal. Sending SIGINT enables Snort to shut down gracefully, subsequently producing statistics for the period measured. This includes metrics like incoming throughput, the count of analyzed packets, and any outstanding packets, if present. Each experiment was conducted three times, and the average value was calculated across these three runs.

\subsection{Results and Discussion}

\begin{figure*}[ht]
    \centering
    \begin{minipage}{0.24\textwidth}
        \centering
        \includegraphics[width=1\linewidth]{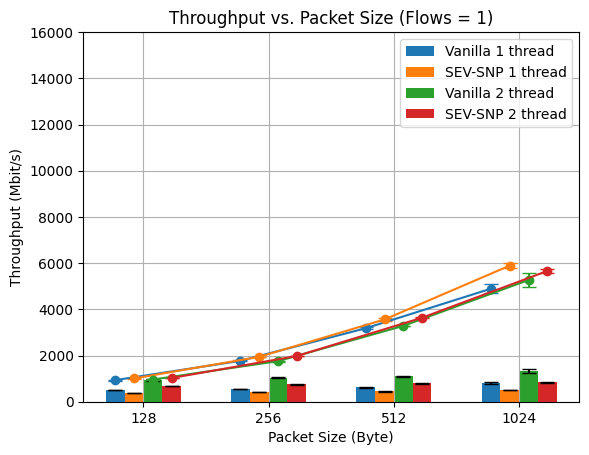}
        \subcaption{\footnotesize Flow=1}
        \label{fig:tcp_f1} 
    \end{minipage}\hfill
    \begin{minipage}{0.24\textwidth}
        \centering
        \includegraphics[width=1\linewidth]{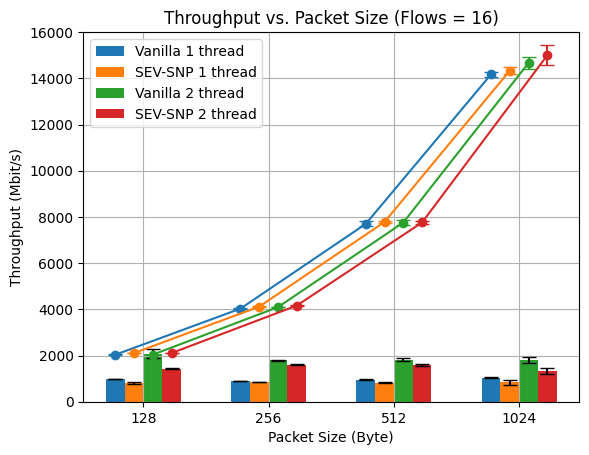}
        \subcaption{\footnotesize Flows=16}
        \label{fig:tcp_f16} 
    \end{minipage}\hfill
    \begin{minipage}{0.24\textwidth}
        \centering
        \includegraphics[width=1\linewidth]{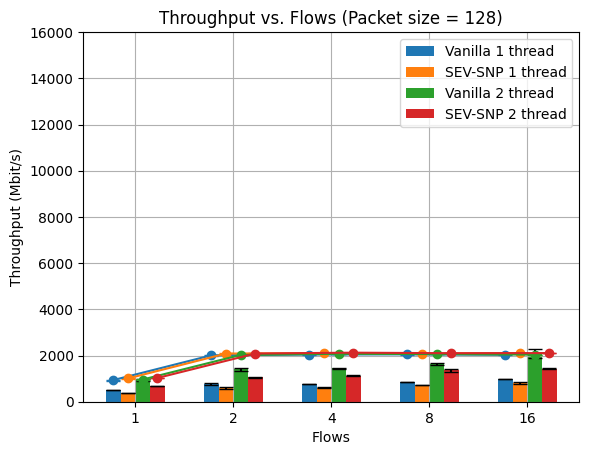}
        \subcaption{\footnotesize Packet size=128}
        \label{fig:tcp_p128} 
    \end{minipage}\hfill
    \begin{minipage}{0.24\textwidth}
        \centering
        \includegraphics[width=1\linewidth]{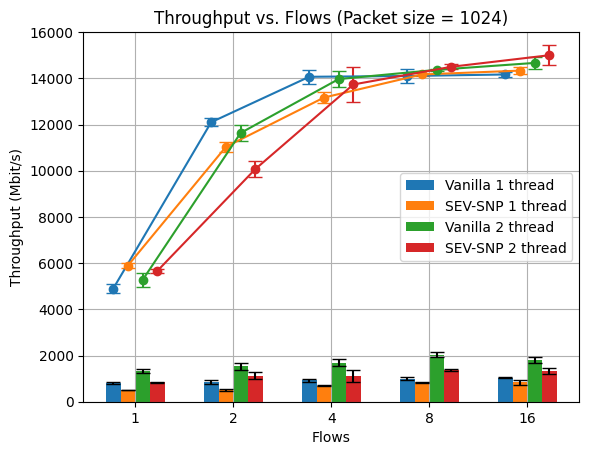}
        \subcaption{\footnotesize Packet size=1024}
        \label{fig:tcp_p1024} 
    \end{minipage} 
    \caption{Throughput of SEV-Snort and Vanilla-Snort with increasing packet size and TCP flows}
    \label{fig:tcp}
\end{figure*}

\begin{figure*}[ht]
    \centering
    \begin{minipage}{0.24\textwidth}
        \centering
        \includegraphics[width=1\linewidth]{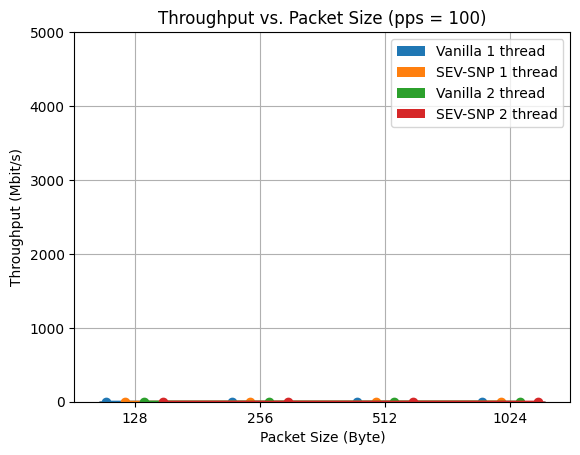}
        \subcaption{\footnotesize 100 pps}
        \label{fig:udp_pps100} 
    \end{minipage}\hfill
    \begin{minipage}{0.24\textwidth}
        \centering
        \includegraphics[width=1\linewidth]{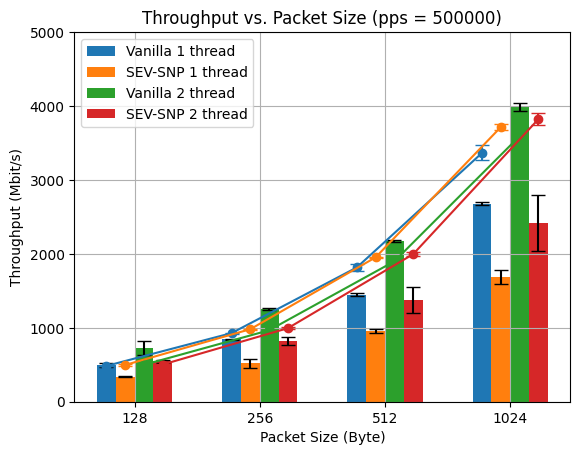}
        \subcaption{\footnotesize 500000 pps}
        \label{fig:udp_pps500k} 
    \end{minipage}\hfill
    \begin{minipage}{0.24\textwidth}
        \centering
        \includegraphics[width=1\linewidth]{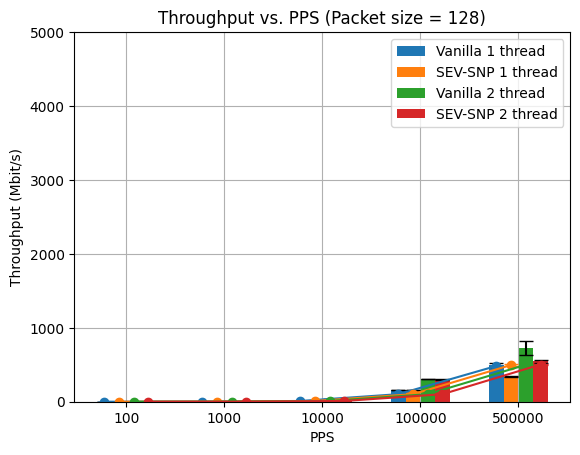}
        \subcaption{\footnotesize Packet size=128}
        \label{fig:udp_p128} 
    \end{minipage}\hfill
    \begin{minipage}{0.24\textwidth}
        \centering
        \includegraphics[width=1\linewidth]{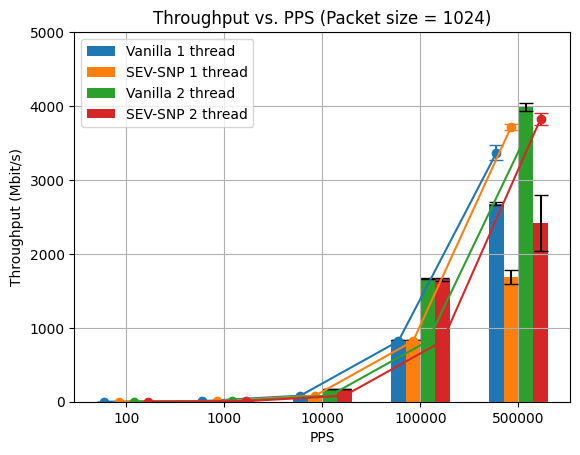}
        \subcaption{\footnotesize Packet size=1024}
        \label{fig:udp_p1024} 
    \end{minipage}    
    \caption{Throughput of SEV-Snort and Vanilla-Snort with increasing packet size and packet per second in UDP}
    \label{fig:udp}
\end{figure*}

In figures \ref{fig:tcp}-\ref{fig:udp}, the bars represent the throughput on the VM's side, sourced from Snort's statistics, while the lines depict the traffic generated from iPerf2 in the host machine. These graphics display variations in packet sizes, the number of TCP flows, and PPS rates. For each parameter, we chose two extreme data points. For example, in figures \ref{fig:tcp_f1} and \ref{fig:tcp_f16}, the X-axis represents packet sizes under conditions of two extreme points of TCP flows. Similarly, in figures \ref{fig:tcp_p128} and \ref{fig:tcp_p1024}, the X-axis indicates the number of TCP flows depicted under two distinct packet sizes at opposing extremes. Also, each figure is represented by two distinct measurements where Snort is run using single thread or dual threads. 

In both TCP and UDP scenarios, there's a clear relationship between packet size and throughput. Specifically, for TCP traffic under varying flows (as seen in figures \ref{fig:tcp_f1} and \ref{fig:tcp_f16}), increasing the packet size directly increases the throughput. This is consistent with the general understanding that larger packets can carry more data per transmission, leading to higher throughput, given the overhead remains relatively constant.

In the UDP scenario, the observations reflect the inherent characteristics of the UDP protocol itself. As depicted in figure \ref{fig:udp_pps100} and \ref{fig:udp_pps500k}, there's a clear uptrend in throughput with the increase in packet size. Unlike TCP, which requires a handshake and acknowledgement mechanism, UDP directly sends packets without prior communication setup. This lack of initial overhead allows UDP to transmit data faster. When observing the throughput with a constant packet size but varying PPS (figure \ref{fig:udp_p128} and \ref{fig:udp_p1024}), an increase in PPS also leads to increased throughput. 

Another immediate observation is that regardless of the volume of generated traffic, Snort caps at a maximum throughput of 2 Gbit/sec for TCP. This peak is reached with two Snort threads and the largest packet size of 1024 bytes when using the vanilla Snort. In contrast, for UDP, Snort matches generated traffic up to 4 Gbit/sec when the packet size is at its maximum. The inherent nature of UDP (handshake-less connection) allows higher caps compared to TCP. 

\subsubsection{On the performance penalty}
\label{penalty}
\begin{figure}[ht]
    \centering
    \includegraphics[width=0.6\linewidth]{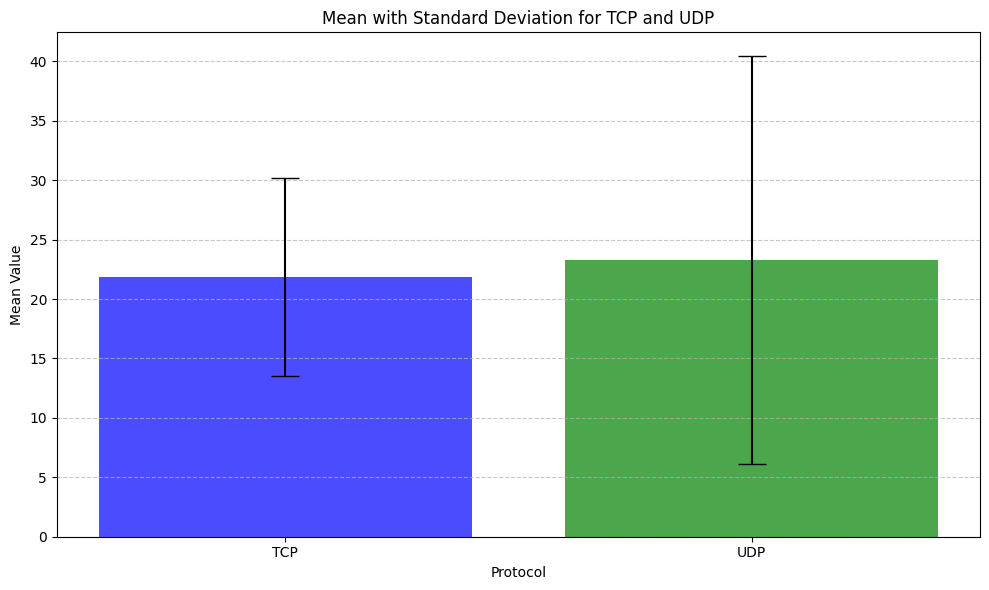}
    \caption{Average performance penalty}
    \label{fig:meanperf}
\end{figure}

Figure \ref{fig:meanperf} presents the average performance penalty observed in both TCP and UDP measurements. This is measured by calculating the average of all the data from the previous measurement in figure \ref{fig:tcp} and \ref{fig:udp}, but not just from the opposite extreme points. It's worth noting that for UDP, the performance penalty is not markedly different for any rate of incoming PPS below 100000, meaning that for these pps rates, SEV does not cause any performance degradation compared to the vanilla Snort; thus, we have excluded calculations from these ranges for clearer results. Such behaviour is anticipated for UDP, given that it doesn't require a handshake for its connections. TCP, on the other hand, behaves differently. Notable performance penalties can be discerned even with small packet sizes or a minimal number of TCP flows. It shows that an average performance penalty is 21.9\% for TCP and 23.3\% for UDP.

\subsubsection{On the comparison with SEC-IDS}
Performance-wise, it is expected that SEC-IDS performs better than SEV-IDS due to the execution location of the Snort; it is the baremetal for SEC-IDS versus in the VM for SEV-IDS. As can be seen from SEC-IDS measurement \cite{snort-sgx}, it can reach near-native performance (close to 100\% throughput in comparison to the vanilla snort). SEV-IDS only has 80\% of the performance compared to the vanilla, depending on the traffic type and load variations, as explained in the previous section. On the other hand, this price comes with its advantages, i.e., no adaptations are needed to make Snort runs off the shelf, as summarized in table \ref{comptab}. 

\section{Related Work}
\label{sec_rw}

The evolution of secure computing and network processing has attracted significant attention in both industry and academia. We focus our related work on the secure deployment of middleboxes and network functions, emphasizing their integration with emerging secure hardware and their performance implications.

Secure Middleboxes with SGX: SGX-Box \cite{10.1145/3106989.3106994} introduced a secure middlebox system design to allow visibility into encrypted traffic without compromising security by leveraging Intel's SGX technology. Similarly, LightBox \cite{10.1145/3319535.3339814} proposed a system for off-site software middleboxes to operate securely and at near-native speeds using Intel SGX. ShieldBox \cite{10.1145/3185467.3185469}, on the other hand, focused on a framework for deploying high-performance network functions over untrusted servers, also relying on Intel SGX. These works underscore the significance of hardware-assisted security in achieving both performance and security for middlebox functions. Our work on SEV-IDS extends this line of investigation by considering AMD's SEV in the context of NFV and analyzing its performance implications.

Decentralized Middlebox Solutions: EndBOX \cite{8416500} presented a system for securely executing middlebox functions on client machines at the network edge, combining VPN with middlebox functions protected by SGX. In a different approach, SafeBricks \cite{safebrickspoddar2018} proposed shielding network functions from an untrusted cloud, ensuring that only encrypted traffic is exposed to the cloud provider, thereby decentralizing the middlebox functionality without compromising on security.

SEV Performance: Our SEV-IDS work builds on this rich landscape of research by delving deep into the performance penalties associated with AMD SEV when deployed in NFV infrastructures. We provide an initial analysis of the trade-offs and benefits of using SEV, extending the discourse on the secure deployment of middleboxes and network functions.

\section{Conclusions}
This paper presents our firsthand experience with running an NFV application inside AMD SEV. Our primary aim is to assess the performance implications of deploying such an application within a trusted VM. Drawing inspiration from similar efforts with Intel SGX, we undertook a comparative analysis. Our findings reveal distinct advantages with AMD SEV. Unlike its SGX counterpart, deploying the NF application within the trusted VM is considerably more straightforward. Specifically, there's no need to modify the NF codebase, manage packet processing (provided a native packet processor is employed), or adjust the trusted clock, as was necessary with SEC-IDS.

Performance-wise, both TCP and UDP workloads exhibited a performance penalty of approximately 20\% when using AMD SEV compared to a non-SEV environment, that is easy to coupe it by assigning more resources. The implications of this penalty, in light of the ease of deployment and other benefits, warrant further exploration and might be acceptable for many practical applications, particularly when the application needs to be secured inside an enclave.

\label{sec_conclusion}

\bibliographystyle{IEEEtran}
\bibliography{references}
\end{document}